# Novel Approach for Cybersecurity Workforce Development: A Course in Secure Design

Filipo Sharevski
College of Computing and Digital Media
School of Computing
DePaul University
Chicago, IL, USA
fsharevs@cdm.depaul.edu

Adam Trowbridge
College of Computing and Digital Media
School of Design
DePaul University
Chicago, IL, USA
atrowbr1@cdm.depaul.edu

Jessica Westbrook
College of Computing and Digital Media
School of Design
DePaul University
Chicago, IL, USA
jwestbro@cdm.depaul.edu

*Abstract*— Training the future cybersecurity workforce to respond to emerging threats requires introduction of novel educational interventions into the cybersecurity curriculum. To be effective, these interventions have to incorporate trending knowledge from cybersecurity and other related domains while allowing for experiential learning through hands-on experimentation. To date, the traditional interdisciplinary approach for cybersecurity training has infused political science, law, economics or linguistics knowledge into the cybersecurity curriculum, allowing for limited experimentation. Cybersecurity students were left with little opportunity to acquire knowledge, skills, and abilities in domains outside of these. Also, students in outside majors had no options to get into cybersecurity. With this in mind, we developed an interdisciplinary course for experiential learning in the fields of cybersecurity and interaction design. The inaugural course teaches students from cybersecurity, user interaction design, and visual design the principles of designing for secure use – or *secure design* – and allows them to apply them for prototyping of Internet-of-Things (IoT) products for smart homes. This paper elaborates on the concepts of secure design and how our approach enhances the training of the future cybersecurity workforce.

## I. Introduction

Engineers, developers, and business people are trained to solve problems. Designers are trained to discover the very problems that need to be solved [1]. Current cybersecurity curricula are focused primarily on responding to existing problems. Engineers learn secure system development life cycle to eliminate security flaws in complex systems [2], [3]. Developers learn secure coding to eliminate software vulnerabilities [4], [5]. Business people learn information security management so companies can address the cybersecurity risk pertaining to their mission critical assets [6], [7]. Very little attention is devoted to training cybersecurity workers on discovering real problems. Designers currently have no training opportunities to build skills in proactively discovering cybersecurity problems.

A common consensus in the academic and professional cybersecurity community is that humans are the number one "problem-makers" in the security of systems [8]. As *legitimate* users, humans settle for the least secure practices: never changing the default password, using easy-to-remember passwords for a multitude of systems, or trusting every email attachment and link, including phishing scams. As *malicious* users, humans never stop looking for ways to exploit systems. From a cybersecurity chain perspective, the real problems are: (i) *insecure use*, which leads to many security updates, and (ii) *intentional misuse,* which leads to continuous security patches. Both problems are user-centered; the insecure use results from security fatigue and user ignorance of security best practice, while the intentional misuse results from security-deficient systems and human desire to violate software or hardware under their control [9], [10].

The conventional approach to solving user-center problems is to train cybersecurity workers in the concepts of *usable security* or how to "deliver required levels of security in systems, software, and hardware while in the same time deliver user effectiveness, efficiency, and satisfaction" [11]. Successful solutions like graphical authentication, password vaults, visual browser cues, and phishing security warnings have emerged, but they neither completely solved the problem of insecure use, nor deterred intentional misuse. In 2017, 81% of all reported data breaches leveraged stolen and/or weak passwords, and 43% of them were obtained by social engineering attacks [12].

Our novel approach to cybersecurity training brings future cybersecurity and user-centered design workers together in discovering and solving cybersecurity problems resulting from insecure use. The inaugural undergraduate-level course in secure design is novel in that it: (1) introduces a new cross-disciplinary field of study between cybersecurity and user-centered design; (2) enables cybersecurity students to proactively discover user-centered problems through interaction design; (3) allows design students to understand the cybersecurity problems through insecure use or intentional misuse experimentation; and (4) encourages collaboration between cybersecurity and design students in secure user-centered design prototyping. This paper introduces the cross-discipline between the cybersecurity and user-centered design – named *secure design*. It also demonstrates how interdisciplinary knowledge was employed to create new training opportunities for future cybersecurity workers while broadening the workforce to include cybersecurity-aware interaction designers.

This project was sponsored by the National Security Agency (NSA) grant for Cybersecurity Curriculum Development GS-17-0003.

## II. Cybersecurity and User-Centered Design

Secure design is a cross-discipline between cybersecurity and user-centered design. Cybersecurity addresses the technology and practices used to protect computer, networking, and cyber-physical systems from harm. User-centered design creates meaningful relationships between humans and technology. Secure design creates meaningful yet secure relationships between humans and technologies that minimizes the risk of harm resulting from insecure use.

Cybersecurity and user-centered design are both optimal when users are able to work uninterrupted. Users are focused on getting things done. They rely on user-centered design to provide a clear path to do so, and they rely on cybersecurity to protect their virtual work environment. Increasing security of a system potentially increases the complexity of getting things done, and pushes users to take short cuts so they can continue working (e.g. passwords on sticky notes on the side of a monitor). Some user-centered design approaches decrease security (e.g. default settings and passwords), bypassing the complexity of secure setup, leaving users vulnerable to very basic attacks. It is important for user-centered design and cybersecurity to converge around a secure design approach, not merely making existing security approaches more "user friendly" but integrating security and usability in such a way that it empowers users. Secure design approaches cybersecurity and user-centered design as foundations for a new approach to designing interactive systems.

The conventional usable security follows the design model shown in Fig 1. There are two concerns captured in the model: (i) the *user concern* of how to make any computer related decision with minimum distraction and (ii) the *security concern* of how to make sure users comply with minimum security. Usable security solutions underestimate the user concern and focus on the security concern to force a supposedly secure decision via user interface, for example enforcing a strong password policy and/or suggesting strong, but random character passwords. Unfortunately, security concerns can be seen as a task-based distraction for users. They usually opt to route around non-task-based blocks and write down all the passwords on a post-it note on their monitors or use weak passwords for all login interfaces.

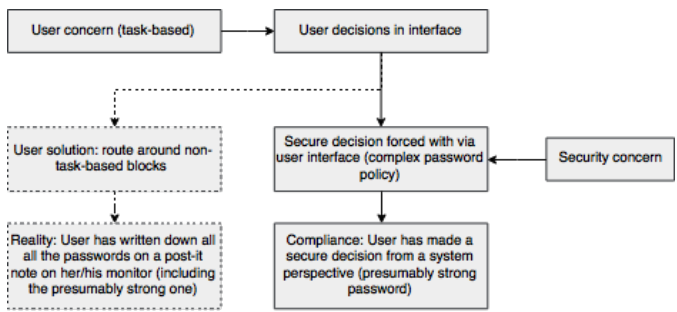

Fig. 1. Usable security design model.

Fig 2. extends the usable security model to a secure design model in which the interface design addresses both the user concerns and the security concerns. The secure design looks into usability design research in order to understand the ergonomics of human-computer interaction. For example, how to apply the principles of interaction design (affordances, signifiers, feedback, constraints, and conceptual models) in creating a secure yet meaningful login interface from the user's perspective.

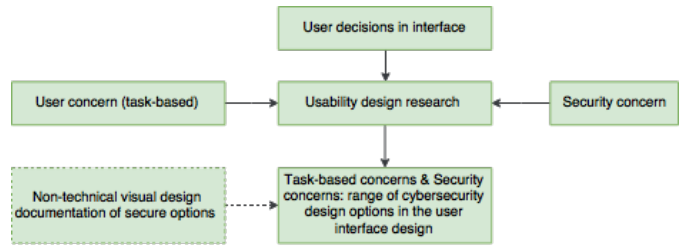

Fig. 2. Secure design model.

*Affordances* refer to the relationships between any object (in this case login prompts) and users. A login prompt affords (is for) system identification and authentication, therefore, it affords access to restricted resources. This is true only if the correct combination of username and password is used. In any other case, login prompts anti-afford or deny access. In short, affordances determine what actions are possible [1]. *Signifiers* communicate how the action should take place. Such an example would be a login prompt highlighting the text "username" and "password" in the respective boxes where these values shall be entered as shown in Fig 3 [13]. *Feedback* is communicating the results of the action. Login prompts display warning icons if either the username or password is missing, or if the combination of both is invalid as shown in Fig. 4a and Fig.4b, respectively. Constraints communicate which actions are permissible. While one can afford to access restricted system resources, there are constraints that pertain to how the username and passwords can be formatted. This is where security concern meets user concern in the human-computer interaction.

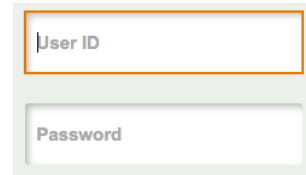

Fig. 3. Login prompt signifiers for username and password.

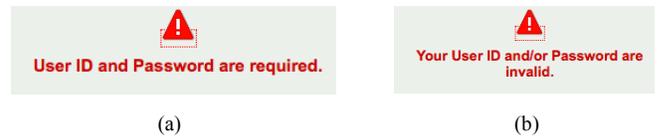

(a)  (b)

Fig. 4. Feedback for (a) missing username/password; (b) invalid username/password.

When communicating the constraints, the usable security approach expects that the user knows exactly how the system works, or has already formed an exact *conceptual model* of identification and authentication. Users are forced to adhere to complex password policies such as one shown in Fig. 5 with no explanation of why they "must" use passwords between 8 and 32 characters, digits, some but not all special characters (perhaps also, why they cannot use different language layouts on a keyboard) [13]. Some login prompts display indicators for "weak" and "strong" passwords that comply with the complex password policy, even though in some cases this can be misleading. For example, the common combination of "Password!123" complies with the indicators for a strong password shown in Fig. 6, yet is still insecure because it is one of the most used default passwords [14]. In addition to these constraints, users are also left to wonder why their workstations block them after three consecutive invalid logins, some websites after five, and some websites after arbitrary number of attempts with invalid username/password combination (protection from online password guessing).

![Fig. 5 password policy rules]

Fig. 5. An example of a complex password policy.

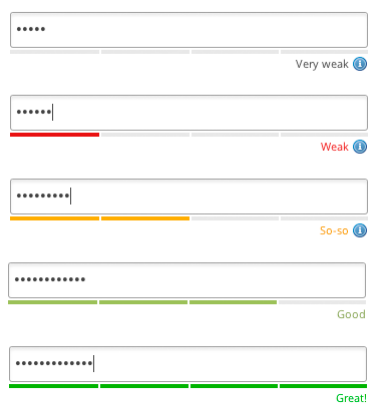

Fig. 6. An example of password strength indicators.

The secure design has no such expectations (for example that users have read the NIST guidelines for electronic authentication and they understand the concept of password entropy [15]); instead it considers that there might be various different conceptual models that are meaningful to different users. The objective is then to create communication -non-technical visual documentation, meaningful signifiers, user understandable feedback, user understandable constraints and justification thereof – to offer options for strong yet meaningful passwords for users. Another objective of secure design is to communicate secure use in a way that enable users to converge on a consolidated and generally accepted conceptual model of identification and authentication. An example of a secure design for guiding users to understand, choose, and form a conceptual model of a strong password is shown in Fig. 7.

### III. A COURSE IN SECURE DESIGN

#### A. Course Description

Secure design is an undergraduate, 300-level experiential learning course for students in cybersecurity, user interaction design, and visual design majors. It involves *classroom instruction*, *hands-on labs*, and *prototyping*. To gain the fundamental knowledge of secure design, classroom instruction is focused on teaching the principles of cybersecurity, user interaction design, and visual design. To apply this knowledge, students work on hands-on laboratory exercises focused on cybersecurity and design measures currently implemented in popular Internet-of-Things (IoT) smart home products like smart light bulbs, locks, thermostats, baby monitors, voice assistants, and cameras. The IoT smart home environment was chosen because: (i) it is popular among younger consumers, (ii) it is highly interactive and interesting to use, (iii) IoT is an emerging technology yet to mature in secure design. The main deliverable in the course is a secure design prototype of an IoT device that incorporates both the user concerns and security concerns while retaining the ergonomics of human-IoT interaction for everyday use.

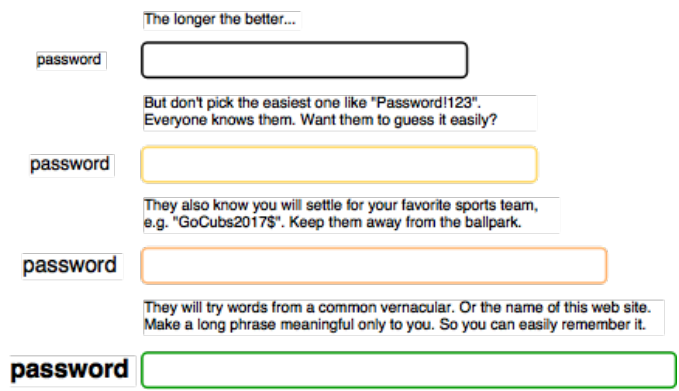

Fig. 7. An example of a secure design for password strength indicators.

#### B. Learning Outcomes

The learning outcomes for the class are classified in two categories according to Bloom's taxonomy [16]. These categories correspond to the knowledge, skills, and abilities acquired out of a course, that is, after the completion of the secure design course students will be able to:

*1) Remember and Understand*

- The cybersecurity principles of confidentiality, integrity, availability
- The general system security mechanisms of identification & authentication, access control, and accountability
- The user interaction design principles: discoverability, affordance, signifiers, feedback, constraint
- The visual design concepts of alignment, balance, order, contrast, proximity & similarity, rhythm, and whitespace

With the ability to recall these principles and explain the concepts of cybersecurity and user-centered design, students will be able to recognize the security flaws in the structure, user interaction, and visual design that enable for insecure use of the current smart home IoT products.

*2) Apply and Analyze*

- Insecure usability issues in smart home IoT products using usability heuristics for secure user interface design

- Security flaws in a smart home IoT application logic, including those introduced by the technology and those that could be introduced by an end user

Capable of drawing connections among the ideas of security and user-centered design, students are trained to discover secure usability issues and validate in smart home IoT interface designs. The ability to discover a cybersecurity problem that impedes the usability enables students to try and solving by prototyping a secure smart home IoT application that eliminates security design flaws.

*C. Course Assessement*

The acquired expertise in secure design is evaluated using several classroom instruments, including *formative* and *summative* assessments. For the "remember and understand" category, discussions and tests were developed to help guide the classroom instruction and measure student competency. Faculty or instructors use the formative conversations to understand which user interaction or visual design concepts need more attention from the cybersecurity students, and which cybersecurity concepts need more attention from user interaction and visual design students. The tests serves as a summative assessment to find out the knowledge, skills, and abilities (from Table 1, Table 2, and Table 3 in the next section) developed by each student for a particular course module.

For the "apply and analyze" category, student reflection writing is selected where students need to identify and describe, in a specific and detailed manner, how the principles of secure design apply to a real-world example of a smart home IoT product. The student reflection writing is formatively assessed to understand whether user interaction and visual design students were able to capture and articulate any technology-introduced flaws in this IoT product, and whether cybersecurity students were able to capture and articulate flaws introduced by the product design.

For the final assignment, the secure design prototype, students are asked to elaborate, in a specific and detailed manner, their prototype and how they have addressed both the usability and security concerns with their choice of a secure design. Faculty and instructors receive rubrics for summative assessment to ensure each secure design prototype has addressed all the security flaws previously identified in the home IoT product. As part of the final assignment, each student presents their prototype to the entire class so the other students can write a peer review report that analyzes the secure design of her/his peer's prototypes. The peer review reports are also part of the final summative assessment, ensuring that students developed skills for analyzing future secure designs.

## IV. Secure Design and Cybersecurity Workforce Development

The challenge of cybersecurity workforce training is to overcome the shortage of skilled cybersecurity workers and educate them to respond to emerging threats [17]. We recognize that the current cybersecurity education centers on technical, management, and policy programs. Students from outside majors have few opportunities to learn and take interdisciplinary cybersecurity courses. A skilled cybersecurity workforce includes not only technically, politically, or economically trained professionals, but also workers who apply novel interdisciplinary knowledge from cybersecurity and user-centered design. By educating interaction and visual design students in the concepts of cybersecurity, knowledgeable and skilled professionals will be added to the future cybersecurity workforce, able to support users in making more secure decisions. By educating cybersecurity students in interaction design and security communication, the future workers will understand the synergy between security and usability and incorporate it into their work tasks.

With this in mind, we have identified the roles, knowledge areas, skills, and abilities related to the secure design course according to the National Cybersecurity Workforce Framework (NCWF) [17]. In addition, we propose several new roles, knowledge areas, skills, and abilities to accommodate the future cybersecurity professionals trained in secure design.

*A. Roles*

From the *Securely Provision (SP)* area of interest, we have identified the following cybersecurity workforce roles for which the secure design students are well-suited:

- Secure Software Assessor (SP-DEV-002)
- Enterprise Architect – (SP-ARC-001)
- Security Architect – (SP-ARC-002)
- Systems Requirements Planner – (SP-RP-001)
- Systems Developer - SP-SYS-002

However, these roles are too general for secure design professionals. We suggest the *Securely Provision (SP)* category to include a *Secure Design* specialty area that has the following three new work roles related to the secure design course:

- <u>Secure Design Architect</u> - "Designs systems security throughout the development life-cycle; translates technology, human-computer interaction principles, and visual design principles into secure designs and processes; considers users' needs, abilities, how they act, how they understand the cybersecurity threat and their part in reducing threat"

- <u>Secure Interaction Designer</u> – "Designs user-experiences using mockups based on user need and ability, in line with cybersecurity principles; collaborates closely with secure design Architect and Analyst to design usable, end-to-end secure system; performs user research and usability tests, with a focus on usability of security and privacy features."

- <u>Secure Visual Designer</u> - "Designs visual e.g. and communication systems for interactive products, as well as documentation and communication systems outside products; translates complex security and privacy principles for end users, based on user profiles developed in conjunction with Security Interaction Designer; design user interaction with security procedures and decisions from a content perspective."

*B. Knowledge Areas*

We have identified the knowledge areas covered by the secure design course in Table 1.

TABLE I.  NCWF KNOWLEDGE AREAS COVERED IN THE SECURE DESIGN COURSE

| Segment | Reference Numbers |
|---|---|
| Knowledge Areas | K0001, K0002, K0003, K0004, K0005, K0006, K0012, K0018, K0019, K0036, K0045, K0058, K0062, K0085, K0086, K0087, K0090, K0147, K0150 K00161, K0165, K0297, K0229, K0362, K0375 |

Given that our course brings new interdisciplinary knowledge not captured in the NCWF, we propose two new *knowledge areas* to be added for which the students will be trained during the secure design course:

- Knowledge of interaction design methods and how changes in the interaction will affect the overall security of the product
- Knowledge of visual design methods and how changes in the graphics will affect the overall security of the product

*C. Skills*

Table 2 lists the NCWF skills that students will develop by taking the secure design course.

TABLE II.  NCWF SKILLS COVERED IN THE SECURE DESIGN COURSE

| Segment | Reference Numbers |
|---|---|
| Skills | S0001, S0005, S0006, S0022, S0023, S0024, S0036, S0051, S0061, S0066, S0077, S0099, S0116, S0122, S0135, S0141, S0147, S0171, S0333, S0357, S0358 |

The prototyping will allow students to develop specific skills for designing and evaluating user interaction that incorporates the main security principles. These skills also enable students to anticipate future threats from insecure use. We suggest four new *skills* to be added in the NCRF for which the students will be trained during the secure design course:

- Skill in applying security principles in user-centered design
- Skill in applying security principles in visual design
- Skill in evaluating the adequacy of security designs for human computer interaction and user communication
- Skill to anticipate new security threats emerging in both system development life cycle, interaction design, and visual design

*D. Abilities*

Table 3 lists the NCWF abilities that students will develop by taking the secure design course.

TABLE III.  NCWF ABILITIES COVERED IN THE SECURE DESIGN COURSE

| Segment | Reference Numbers |
|---|---|
| Abilities | A0001, A0013, A0015, A0016, A0023, A0024, A0041, A0048, A0049, A0054, A0080, A0089, A0096, A0105 |

The secure design course educates design students to be able to incorporate cybersecurity and cybersecurity students to be able to make systems intuitively usable. We suggest that two new *abilities* to be added in the NCRF for which the students will be trained during the secure design course:

- Ability to develop complex systems and product and incorporating security controls considering the principles of interaction and visual design
- Ability to design interaction and user communication to provide the required level of security without sacrificing usability

The main goal of the cybersecurity education is to "promote interest in cybersecurity by increasing quality and diversity of course offerings and research opportunities" [17]. With the surge of sophisticated cyberattacks, the immediate objective is to "improve the cybersecurity workforce pipeline" to overcome the shortage of skilled cybersecurity workers. The secure design course brings quantitative and qualitative improvements to the cybersecurity workforce pipeline: it increases the number of future cybersecurity workers by including new workforce roles and adds new knowledge areas, skills, and abilities the ones identified in NCWF.

V. DISCUSSION

This course offers numerous benefits to the development of the cybersecurity workforce. Secure design is a new subject that is focused on discovering real cybersecurity problems related to insecure use of modern products. Interdisciplinary in nature, it can be adopted at both traditional cybersecurity programs, but also offered at interaction design and visual design schools. Employing an experiential learning approach, it emphasizes practical learning and skills development through laboratory experimentation, prototyping, and interdisciplinary teamwork on emerging technologies, such as IoT smart home products.

The interdisciplinary approach is not new to cybersecurity education; Authors in [18] identified three pillars of domain consisting of people, process and technology that need to be integrated into cybersecurity education programs. Secure design covers each of these domains. Information design communicates processes and technology to people; Interaction design structures processes for people to use technology; Cybersecurity ensures that this use is not malicious and does not result in harm to the people, their data, or technologies. Usable security is interdisciplinary, but arises from two separate disciplinary approaches. Secure design begins as cybersecurity and user-centered design pedagogy combine into a single design approach in which security and usability become inseparable.

The secure design course can be modified to serve particular service/product that is of interest to the education institution, i.e. software, hardware, cyber-physical systems, etc. (the course material generic and customizable to various other technologies). As a unique hands-on training intervention, the course can also be used in public-private partnerships for a specialized preparation of cybersecurity professionals as well as secure interaction designers.

The secure design course impacts several dimensions of cybersecurity training by introducing a novel and human-centric perspective on how systems and products are conceptualized, designed, and built incorporating cybersecurity. Perhaps most importantly, it allows students from different majors to develop a common cybersecurity and secure design lexicon so the user concerns and security concerns can be better communicated over the entire product development life-cycle. The future cybersecurity workers with specific knowledge, skills, and abilities to secure design will be able to broaden the field of usable security to be more user-centered, emphasizing the importance of communicating security needs to end users to eliminate insecure use.

Secure design extends cybersecurity awareness to the students who will go on to design not only the devices and user interfaces, but also the communication materials (e.g. documentation, signage). Through these materials, users will learn cybersecurity protocol and, most importantly, why some cybersecurity interfaces are necessary, and must become part of "getting things done." This approach is unique, enhances cybersecurity infrastructure for research and education, and offers a design approach beyond existing "usable security" approaches.

The secure design course also impacts the larger community. While the cybersecurity workforce is only 11% women [19], 54% of designers are women, and two-thirds of design students surveyed by AIGA, the professional association for design, were women [20]. We believe that by increasing the scope of cybersecurity education to include interaction and graphic designers we may encourage more women students working with cybersecurity principles and practices, ready to participate in cybersecurity from the design side of development teams.